\RequirePackage{lineno}
\documentclass[12pt,amssymb,showkeys,superscriptaddress,preprint,tightenlines]{revtex4-2}
\usepackage{hyperref}
\usepackage{times}
\usepackage{amsmath}
\usepackage{graphicx}
\usepackage{epsfig}
\usepackage{dcolumn}
\usepackage{bm}
\usepackage{color}
\usepackage{longtable}
\usepackage{array}
\usepackage{setspace}
\usepackage{multirow}
\usepackage{threeparttable}
\usepackage{epstopdf}
\newcommand{\lum}{{\cal L}}
\newcommand{\eff}{\varepsilon}

\newcommand{\psip}{\psi(2S)}

\newcommand{\jpsi}{J/\psi}
\newcommand{\ECM}{E_{\rm CM}}

\newcommand{\EE}{e^+e^-}
\newcommand{\MM}{\mu^+\mu^-}

\newcommand{\oldchange}[1]{\textcolor{black}{#1}} % changes that are not new anymore

\newcommand{\revise}{\textcolor{black}}

\begin{document}

\title{Measurement of integrated luminosities at BESIII for data samples at center-of-mass energies between 4.0 and 4.6 GeV}
\author{\begin{small}
\begin{center}
M.~Ablikim$^{1}$, M.~N.~Achasov$^{10,b}$, P.~Adlarson$^{68}$, S. ~Ahmed$^{14}$, M.~Albrecht$^{4}$, R.~Aliberti$^{28}$, A.~Amoroso$^{67A,67C}$, M.~R.~An$^{32}$, Q.~An$^{64,50}$, X.~H.~Bai$^{58}$
% ...
\\
\vspace{0.2cm}
(BESIII Collaboration)\\
\vspace{0.2cm} {\it
$^{1}$ Institute of High Energy Physics, Beijing 100049, People's Republic of China\\
% ...
$^{4}$ Bochum Ruhr-University, D-44780 Bochum, Germany\\
% ...
$^{10}$ G.I. Budker Institute of Nuclear Physics SB RAS (BINP), Novosibirsk 630090, Russia\\
% ...
$^{14}$ Helmholtz Institute Mainz, Staudinger Weg 18, D-55099 Mainz, Germany\\
% ...
$^{28}$ Johannes Gutenberg University of Mainz, Johann-Joachim-Becher-Weg 45, D-55099 Mainz, Germany\\
% ...
$^{32}$ Liaoning Normal University, Dalian 116029, People's Republic of China\\
% ...
$^{50}$ State Key Laboratory of Particle Detection and Electronics, Beijing 100049, Hefei 230026, People's Republic of China\\
% ...
$^{58}$ University of Jinan, Jinan 250022, People's Republic of China\\
% ...
$^{64}$ University of Science and Technology of China, Hefei 230026, People's Republic of China\\
% ...
$^{67}$ University of Turin and INFN, (A)University of Turin, I-10125, Turin, Italy; (B)University of Eastern Piedmont, I-15121, Alessandria, Italy; (C)INFN, I-10125, Turin, Italy\\
$^{68}$ Uppsala University, Box 516, SE-75120 Uppsala, Sweden\\
% ...
$^{b}$ Also at the Novosibirsk State University, Novosibirsk, 630090, Russia\\
% ...
}
}
\date{\today}

\begin{abstract}

The integrated luminosities of the data samples collected \revise{in the BESIII experiment} in
2016--2017 at center-of-mass energies between 4.19 and 4.28~GeV
are measured with a precision better than 1\% by analyzing
large-angle Bhabha scattering events. The
integrated luminosities of the \revise{old} data sets collected in 2010--2014
are updated by considering correction related
to the detector performance, \revise{offsettting the effect of newly discovered readout errors in the electromagnetic calorimeter that happen haphazardly.}
%These results are essential for precision measurements of the production cross sections of the charmonium and charmonium-like states.
%\revise{The data sets are used to study charmonium-like states and higher excited charmonium states, and they can also be used for analyses of charm physics, R value, and so on. }

\end{abstract}


\maketitle

\section{Introduction}

In recent years, the newly discovered charmonium-like states have drawn
great attention due to their exotic properties~\cite{reviews}. These
states are above the open-charm threshold, and their strong
coupling to hidden-charm processes suggests that they could be candidates for unconventional charmonium states.
Study of the properties of these states, through either verifying or excluding 
possible interpretations about their exotic nature (such as molecular states,
tetraquark states, hybrid states, etc.), or establishing the connection between these states
and higher excited charmonium states, 
has \revise{the} potential to provide more insight into the quark model and a better understanding of quantum chromodynamics (QCD).

The BESIII experiment~\cite{bes3}, which operates at the  $\tau$-charm factory BEPCII~\cite{bepc2}, has collected the world's largest
$\EE$ collision data samples at center-of-mass (CM) energies
between 3.81 and 4.60~GeV~\cite{white}. In this energy region, the
charmonium-like states (also called XYZ states), together
with higher excited charmonium states, can be produced copiously, and comprehensive studies of these particles can be performed.
\revise{Also, the data can be used in other studies beyond the field of charmonium physics, such as R measurement or on various topics in charm physics.}
%\revise{Also, the data can prove the validity of pQCD in the description of R as required by a theory-based evaluation of (?). On the other hand, the data can be directly used as input in the dispersion integral in a data-driven approach. }

To shed light on the topics mentioned above, it is essential to measure the production cross sections of these states,
which in return require the precise knowledge of the time-integrated luminosities of the relevant data samples.

In this paper, we present the results of the luminosity measurements
for the XYZ data samples taken by BESIII from December 2016 to May
2017, as well as an update on the previous measurement for the XYZ
data samples taken from December 2011 to May
2014~\cite{PreviousAnalysis}. The update is necessary 
 since 
%a malfunction of the detector that was not modelled in Monte Carlo (MC) simulation has been recently discovered,
%which resulted in an underestimation of the previously measured integrated luminosities. 
\revise{a malfunction of the detector that was not modelled in Monte Carlo (MC) simulation, which resulted in an underestimation of the previously measured integrated luminosities, has been recently discovered.}
The measurement is based on analysis of the Bhabha scattering process $\EE\to
(\gamma)\EE$,
\revise{and the procedure we take is similar to the one in a previous BESIII analysis~\cite{PreviousAnalysis}. }
The process is chosen for its clean signature and large production cross section, which is known with high theoretical precision. These features allow a
precise measurement with small statistical and systematic
uncertainties.

\section{The BESIII detector and data samples}

BESIII is a general purpose detector which operates at the $e^+e^-$ collider
BEPCII~\cite{bepc2}. Due to the crossing angle of the beams at the
interaction point, the $e^+e^-$ CM system is slightly boosted with
respect to the laboratory frame. A detailed description of the
facility is given in Ref.~\cite{bes3}. The cylindrical core of the
BESIII detector covers 93\% of the full solid angle and consists
of a helium-based multilayer drift chamber~(MDC), a plastic
scintillator time-of-flight system~(TOF), and a CsI(Tl)
electromagnetic calorimeter~(EMC), which are all enclosed in a
superconducting solenoidal magnet providing a 1.0~T magnetic
field. The solenoid is supported by an octagonal flux-return yoke
with resistive plate counter muon identification modules
interleaved with steel. The charged-particle momentum resolution
at $1~{\rm GeV}/c$ is $0.5\%$, and the d$E$/d$x$ resolution is $6\%$
for electrons from Bhabha scattering. The EMC measures photon
energies with a resolution of $2.5\%$ ($5\%$) at $1$~GeV in the
barrel (end-cap) region. The time resolution in the TOF barrel
region is 68~ps, whereas that in the end-cap region is 110~ps. The
end-cap TOF system was upgraded in 2015 using multi-gap resistive
plate chamber technology, providing a time resolution of
60~ps~\cite{etof}. A {\sc geant4}~\cite{Allison:2006ve} based
detector simulation package has been developed to model the
detector response.

From December 2016 to May 2017, eight data sets were taken at CM
energies between 4.19 and 4.28~GeV. These data sets were collected in the
vicinity of the $Y(4230)$ and $Y(4320)$ resonances, aimed at
studying the line shapes of the production cross sections and the decay
properties of these charmonium-like states. The CM energy ($\ECM$)
of each data sample has been determined with the process $\EE\to \MM$~\cite{BESIII:2020eyu}, and is listed in Table~\ref{tab:result_new}.

\begin{table}[htbp]
\centering \caption{\label{tab:result_new} Summary of the integrated luminosity results for the 2016--2017 XYZ data
samples. $N_{\rm cor}$ is the number of events \revise{recovered by} the \revise{correction for the EMC readout error} as defined in Section \ref{subsec:emc_correction}. The first uncertainties are statistical and the second ones are systematic.}
\setlength{\tabcolsep}{2mm}{
\begin{tabular}{cccccccc}
\hline
 Sample & $\ECM$ (MeV) & $N_{\rm obs}$ ($\times 10^6$)
        & $N_{\rm cor}$ ($\times 10^6$) & $\sigma_{Bhabha}$ (nb)
        & $\eff$ (\%) & $\lum$ ($\rm pb^{-1}$) \\
\hline
4190 & 4189.12 & 32.62 & 0.04 & 354.82 & 17.60 & 526.7$\pm 0.1 \pm 2.2$\\
4200 & 4199.15 & 32.59 & 0.05 & 353.88 & 17.53 & 526.0$\pm 0.1 \pm 2.1$\\
4210 & 4209.39 & 31.73 & 0.05 & 352.98 & 17.40 & 517.1$\pm 0.1 \pm 1.8$\\
4220 & 4218.93 & 31.45 & 0.05 & 352.42 & 17.41 & 514.6$\pm 0.1 \pm 1.8$\\
4237 & 4235.77 & 32.32 & 0.07 & 350.79 & 17.41 & 530.3$\pm 0.1 \pm 2.7$\\
4246 & 4243.97 & 32.65 & 0.07 & 350.26 & 17.38 & 538.1$\pm 0.1 \pm 2.6$\\
4270 & 4266.81 & 31.86 & 0.08 & 348.01 & 17.31 & 531.1$\pm 0.1 \pm 3.1$\\
4280 & 4277.78 & 10.46 & 0.03 & 346.92 & 17.21 & 175.7$\pm 0.1 \pm 1.0$\\
\hline
\end{tabular}}
\end{table}
%Maybe ECM should be cited together with errors? They are present in [8]
%This table would need a description of all the events, if not cited in the text

For each data set, two million Bhabha events were generated with
the {\sc babayaga@nlo}~\cite{Balossini:2008xr} generator, using the
parameters presented in Table~\ref{tab:nlo_options}. In the
simulation, the scattering polar angle of the final state
electrons has been limited to a range from $20^\circ$ ($\theta_{\rm
min}$) to $160^\circ$ ($\theta_{\rm max}$), fully covering the
detector acceptance. The beam energy is set to the value
determined with $\EE\to \MM$ events in the same data
set~\cite{BESIII:2020eyu}, and the energy spread is set to be 1.364~MeV. An
energy threshold of 1.0~GeV ($E_{\rm min}$) is applied to the
final-state electrons and positrons. 
%The acollinearity of the events, which is defined as the angle between the electron and the
%reverse extension line of the positron, is not constrained, nor is the number of photons from initial/final state radiation. 
The acollinearity of the events (i.e. the angle between the electron and the reverse extension line of the positron) and the number of photons from initial/final state radiation are not constrained.
Additionally, a selection on the invariant mass of
the $e^+e^-$ pair ($M(\EE)$) larger than
3.8~GeV/$c^2$ has been applied, to reduce the computing time for simulation by avoiding the need to sample over narrow states such as the 
$\psip$ and $\jpsi$ resonances.

\begin{table}[htbp]
\centering \caption{\label{tab:nlo_options} Parameters of the {\sc
babayaga@nlo} generator for MC sample
at $\ECM=4.19$~GeV. For the other energy points, only the 
$\ECM$ setting changes.}
\begin{tabular}{c c}
\hline
Parameter & Value \\
\hline
$\ECM$ (MeV) & $\rm{4189.12}$ \\
Beam Energy Spread (MeV) & $\rm{1.364}$ \\
$\theta_{\rm min}$ ($^\circ$) & $\rm{20}$ \\
$\theta_{\rm max}$ ($^\circ$)& $\rm{160}$ \\
Maximum Acollinearity ($^\circ$)  & $\rm{180}$ \\
$E_{\rm min}$ (GeV)  & 1 \\
$M(\EE)$ (GeV/$c^2$) & $> \rm{3.8}$ \\
\hline
\end{tabular}
\end{table}

%-----------------------------------------------------------------------

\section{Event Selection} 
\label{subsec:lum_eve_sec}

Signal Bhabha candidate events are required to have two oppositely
charged good tracks. The good tracks must originate from a
cylindrical volume, 
centered around the interaction point, with a radius of 1~cm perpendicular to the beam axis and a length of $\pm$10 cm along the beam axis.
The polar angle of the
tracks $\theta_{\rm{MDC}}$, measured by the MDC and
boosted to the $\EE$ CM frame, is required to be in the fiducial volume of $|\cos\theta_{\rm{MDC}}|<0.8$.
The deposited energy of each track
in EMC must be larger than $0.37 \times \ECM$, and the
momentum of each track has to be larger than
$0.47 \times \ECM$, to reduce background from di-muon pairs or from the decays of light resonances, respectively. The invariant mass of the track
pair is required to be larger than 3.85~GeV/$c^2$, because only the
events with an invariant mass above 3.8~GeV/$c^2$ are produced in the MC
event generator.
% honestly speaking  I did not get the meaning of the previous sentence
As demonstrated by a similar previous
analysis~\cite{PreviousAnalysis}, the remaining background contribution after
applying these selection criteria is negligible.

Figure~\ref{fig:data_mc_comparison} shows the comparison between
data and MC simulation for the kinematic variables previously discussed. There is a reasonable
agreement in the distributions of all the variables.

\begin{figure}[htbp]
\centering
\includegraphics[width=0.48\textwidth]{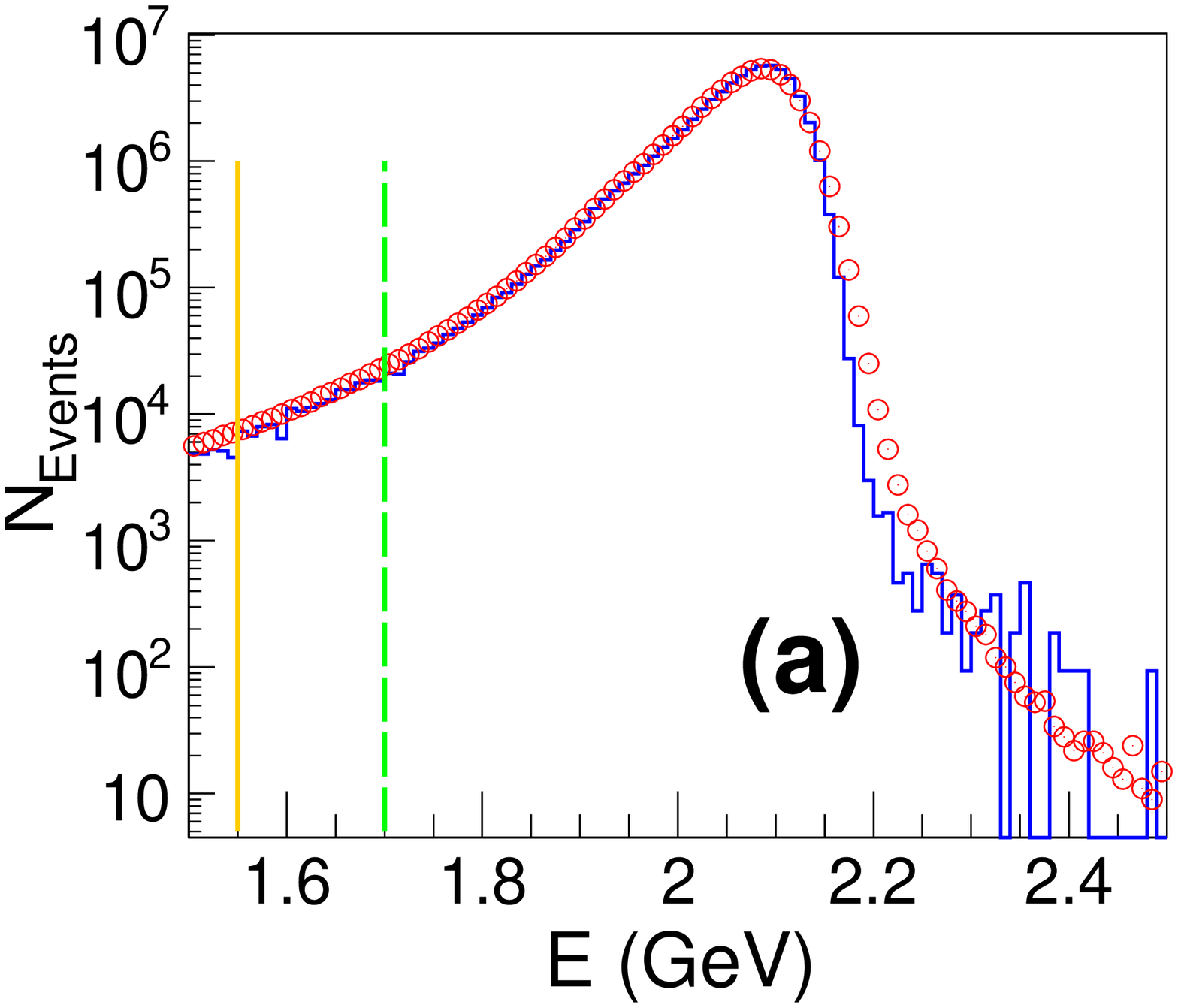}
\includegraphics[width=0.48\textwidth]{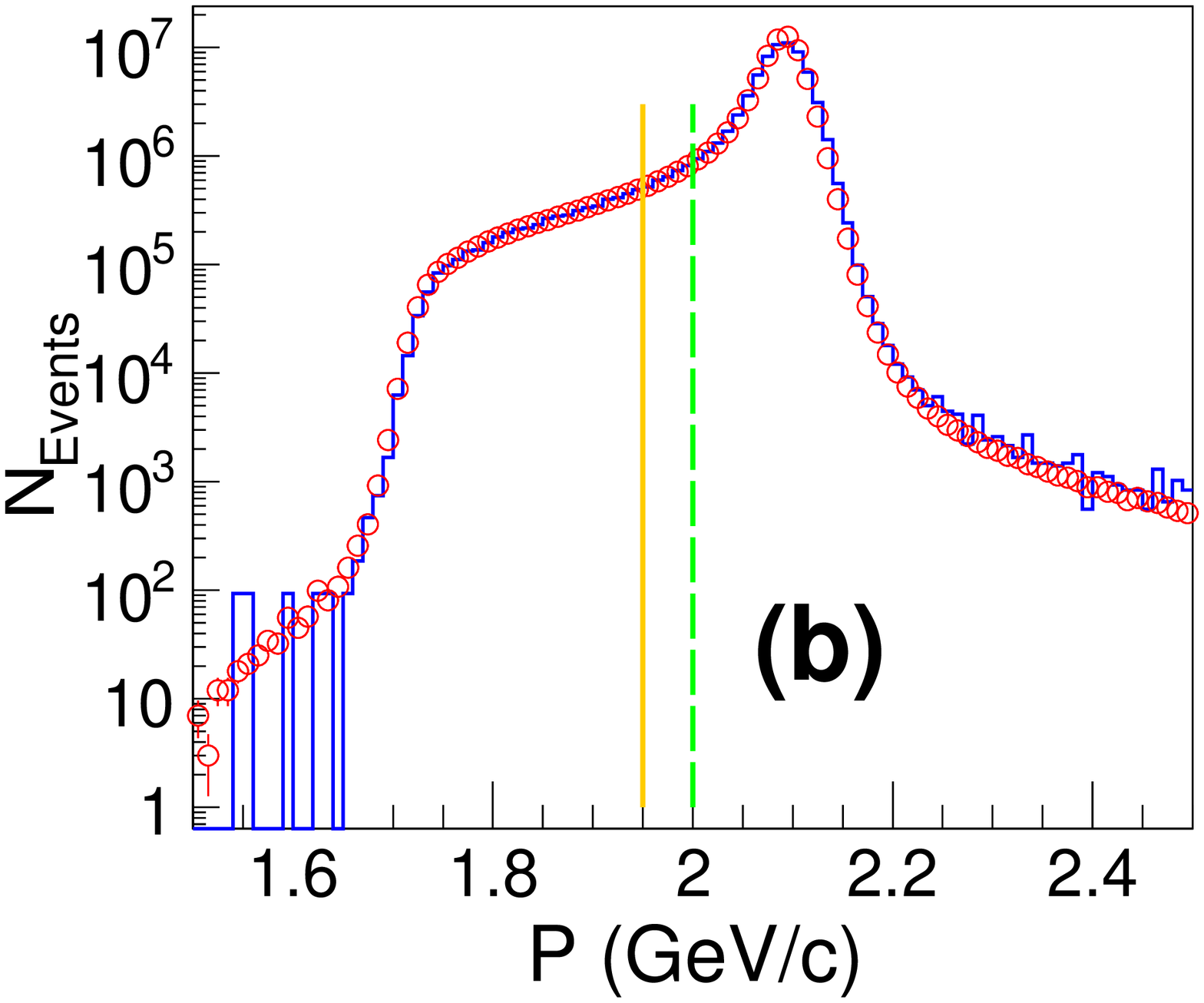} \\
\includegraphics[width=0.48\textwidth]{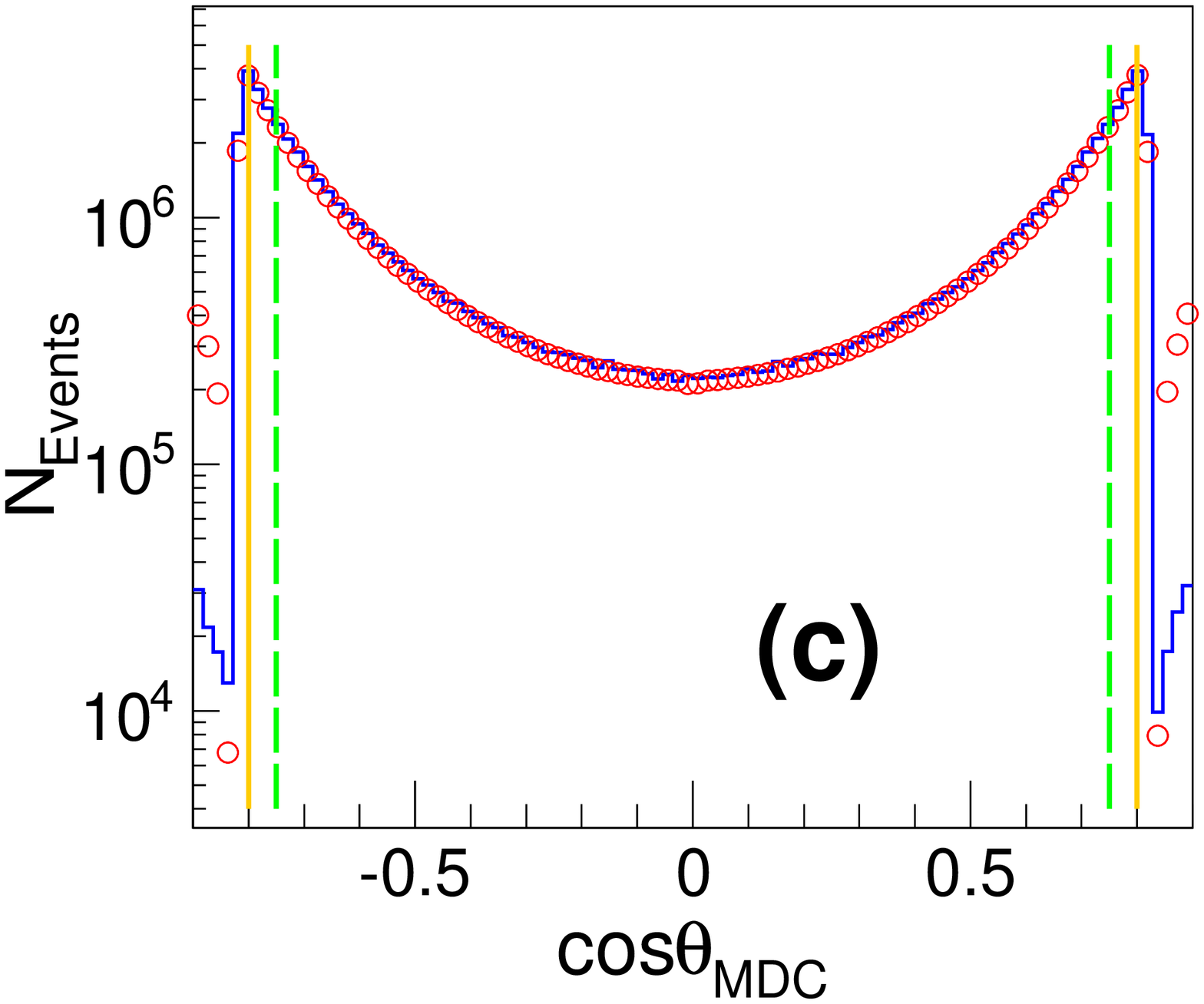}  
\includegraphics[width=0.48\textwidth]{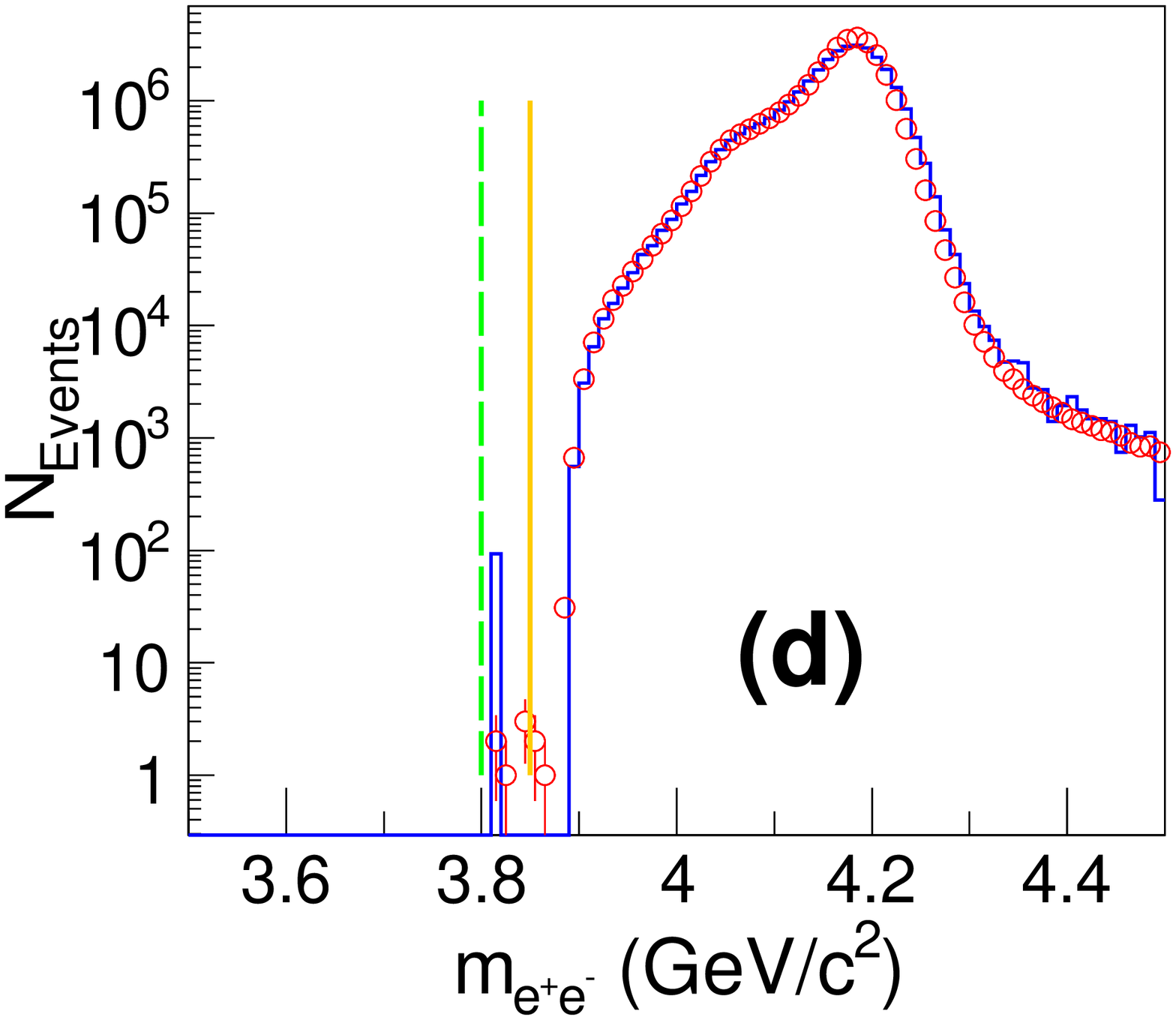}
\caption{\label{fig:data_mc_comparison} Data and MC simulation
comparison for the variables used in the event selection for the
4190 data sample, including the energy deposition in EMC (a),
the momentum (b) and the $\theta$ polar angle (c) of electrons and positrons, as well as the invariant mass of the $e^+e^-$ pair (d). 
%The (a, b, c) plots show the distributions of positrons, and the corresponding distributions for electron tracks are very similar.
Red circles indicate data, while the blue histograms are the MC distributions.
Yellow solid lines mark the thresholds for the standard selection
criteria, while the green dashed lines indicate altered values
used for systematic uncertainty estimation. All the relevant
tracks are boosted to the $\EE$ CM frame.}
\end{figure}

\section{EMC Readout Correction and Luminosity Results} 
\label{subsec:emc_correction}

The energy deposition in the EMC is used to identify the final state
electron/positron tracks. In a study of high energy EMC showers we
found that the EMC electronics occasionally failed to provide valid signals for crystals with 
high deposited energy. 
\revise{The problem mainly occurs for the channels in the absolute polar angle ($\cos{\rm{\theta}}$) ranges  of (0.6, 0.8) close to the horizontal plane.}
To illustrate this issue, consider the left plot of Fig.~\ref{tab:Normal-and-abnormal},
which shows the  EMC energy deposition of a typical high energy (\revise{around} 2 GeV) electron or positron shower, where no problem occurs.  The shower extends across
$5\times5$ crystals and the deposited energy in crystal numbered
(24,~2) is 1592~MeV, one order of magnitude larger than
those of the nearby crystals. In contrast, the right plot of
Fig.~\ref{tab:Normal-and-abnormal}  shows an example of a shower missing the readout of the EMC energy deposition from one crystal. 
Here, the largest energy deposition is expected to be found in the crystal numbered (59,~3) but no valid value is recorded, which leads to an underestimation of the total deposited energy by more than 1~GeV. This effect is not simulated in the MC samples,
and must be corrected.

\begin{figure}[htbp]
\centering

\includegraphics[scale=0.60]{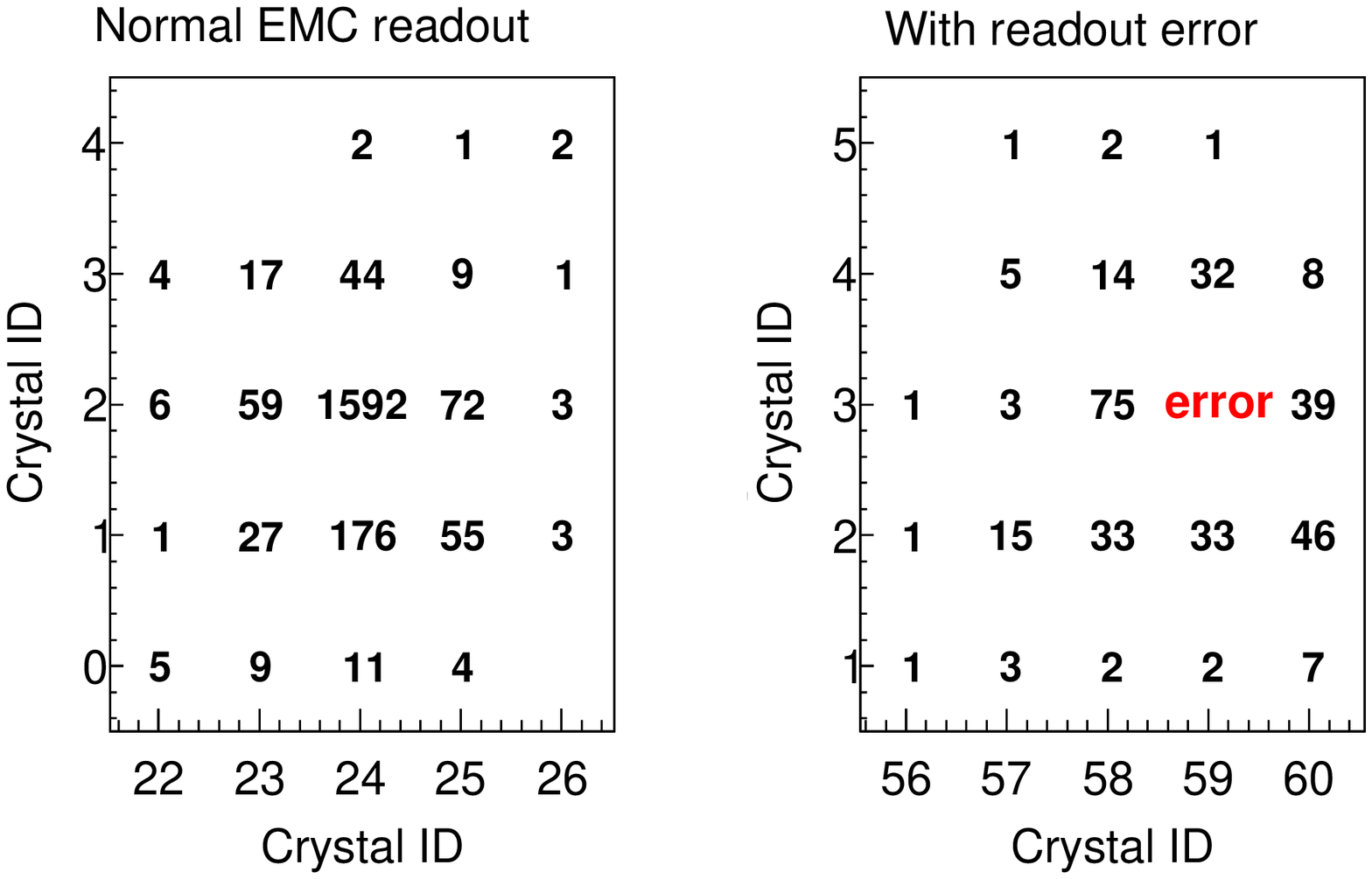}
\caption{\revise{The} EMC energy distributions of a normal EMC shower (left) and an abnormal
one suffering from EMC readout errors (right). The $x$-axis and $y$-axis
mark the EMC crystal ID. The number in each bin represents the
deposited energy in the crystal (in MeV). The hitmap of the
abnormal track is characterized by a missing value in the middle where
major energy deposition is expected.
\label{tab:Normal-and-abnormal}}
\end{figure}

While the reason of this problem is still under investigation,
the amount of affected events can be estimated by searching for the MDC tracks with unexpected EMC
information.
Figure ~\ref{fig:h2ee} shows the dielectron deposited energies for events satisfying all the other requirements of our selection criteria, for the MC sample and the experimental data.
As one can see in the plots, two abnormal peaks can be found in the data samples, which are not present in the MC sample. These peaks are formed by events where the reconstructed energy deposition by the charged track in the EMC is missing the readout signal from one crystal.
To select these events, we apply all the requirements apart from that on the deposited energy in EMC; 
afterwards, we require that the deposited energy in EMC must be larger than $0.37 \times \ECM$
for one track and lower than $0.15 \times \ECM$ for the other. 
The events in which both tracks are affected by \revise{EMC readout errors} are rare, and their contribution is considered as negligible.
Finally, we require that the ionization energy loss of both tracks in MDC must be close to the expected energy loss of electron tracks of the same energy.
%This final piece of requirement is used to increase our confidence about the truth of these self-contradictory events by rejecting some alternative hypotheses (maybe the events aren't consist of $e^+e^-$ pairs, or it's the MDC that malfunctions), and it turns out nearly all events of our interests can pass it.

%Additionally, we require that
%$(1-\rm{normPH}_{\rm p})^2+(1-\rm{normPH}_{\rm m})^2<0.05$, where
%$\rm{normPH}$
% stands for the normalized pulse height of ionization energy lost in MDC.
%As shown in Fig. ~\ref{fig:normph}, the pulse height is normalized to 1 for electrons and to other values for other particles.
%This piece of requirement ensures that the two tracks in such events, despite having large
%differences in their deposited energies in EMC, are indicated by MDC to be good electrons tracks with similar energies. 

Since no normal physics events should be able to pass the above selection, we assume that all the candidates passing the requirements are Bhabha events that suffered from the EMC readout errors.
These events are simply added to the sample of observed Bhabha events whose selection is summarized in  Section \ref{subsec:lum_eve_sec}.

\begin{figure}[htbp]
\centering
\includegraphics[width=0.45\textwidth]{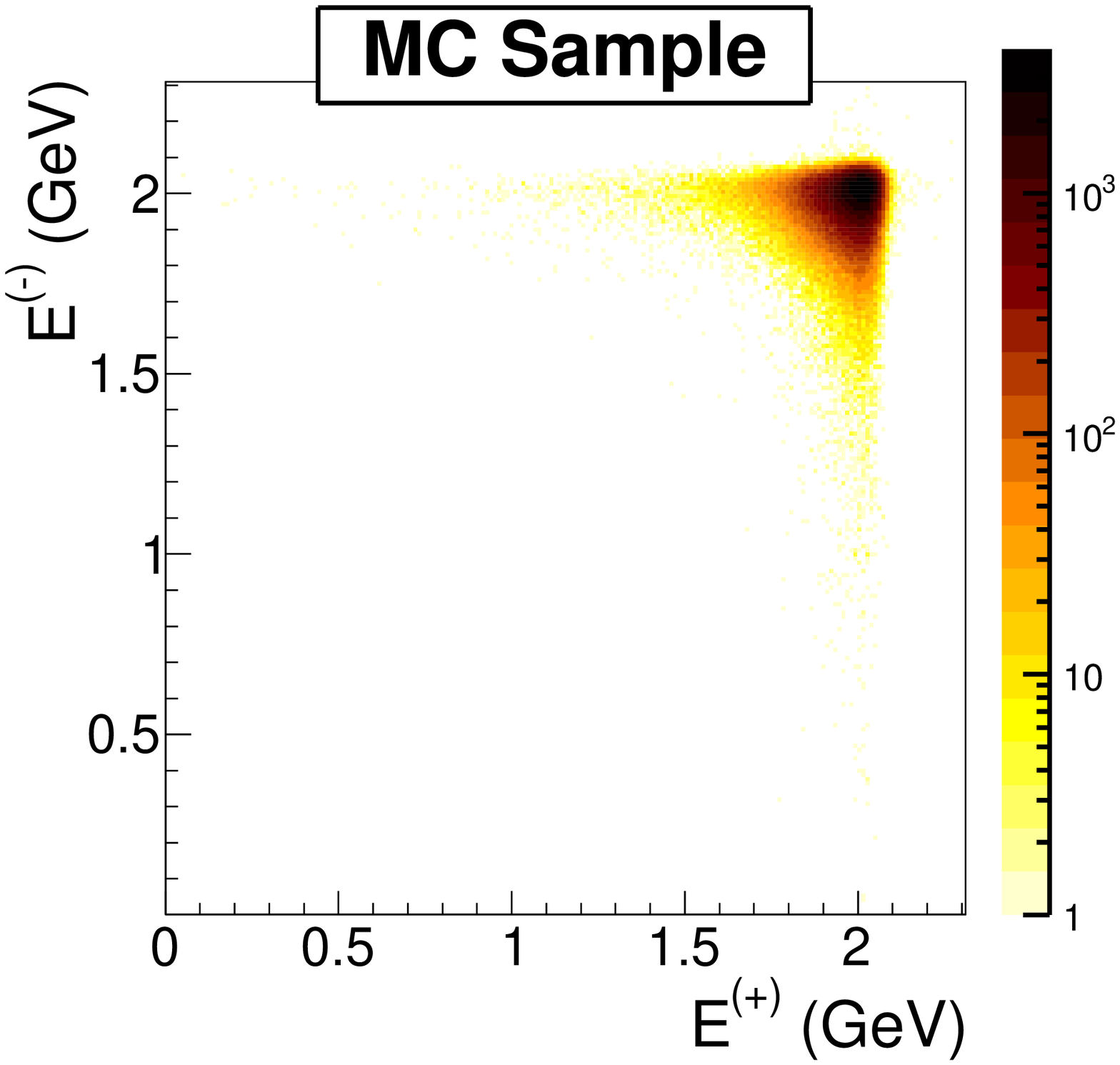}
\includegraphics[width=0.45\textwidth]{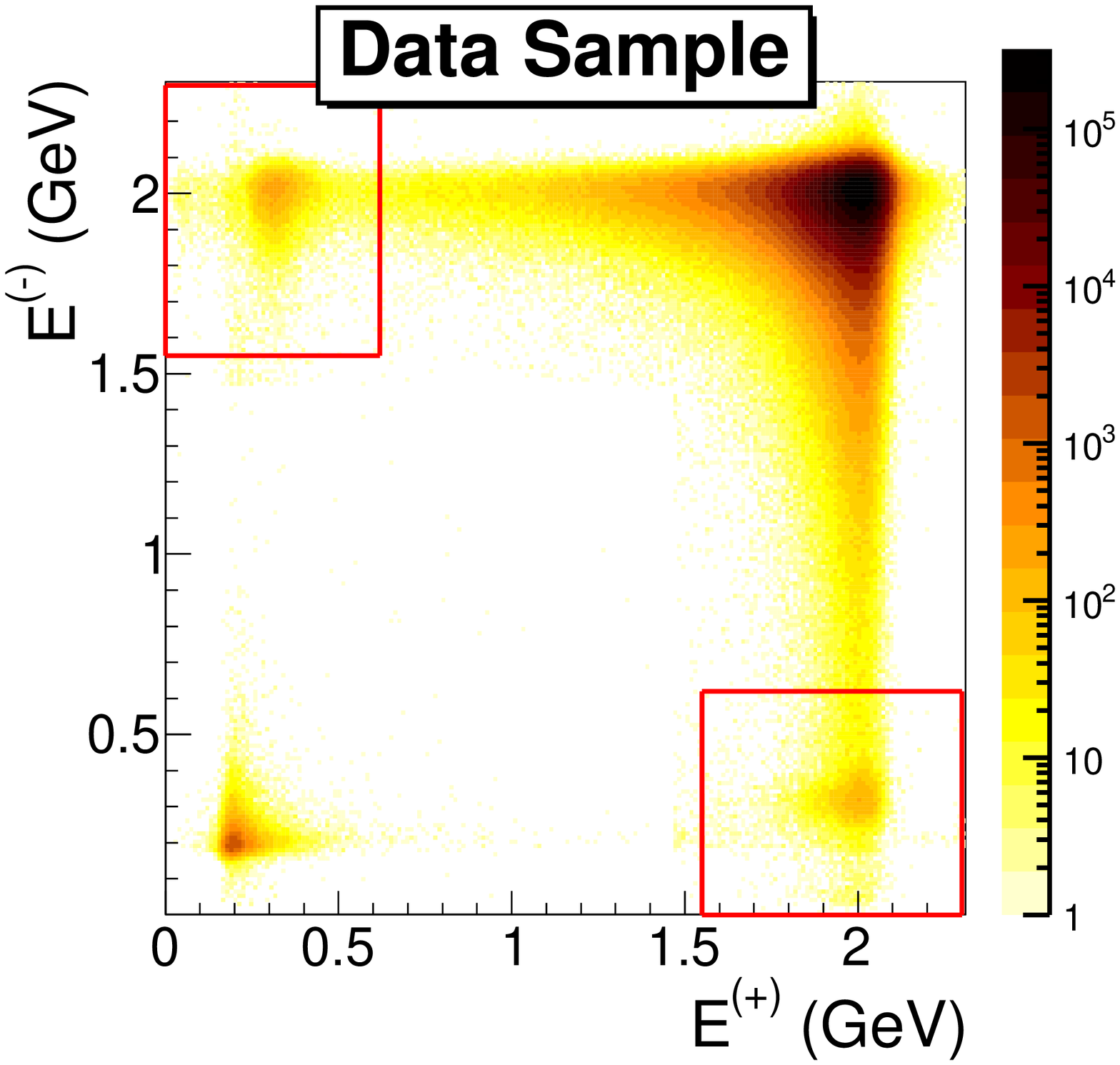}

\caption{
Deposited energies \revise{of positively charged tracks ($\rm{E}^{(+)}$) and negatively charged tracks ($\rm{E}^{(-)}$)} for the events satisfying all requirements except for the EMC energy depositions, for MC sample (left) and data sample (right) at 4.19 GeV. There are large differences between the energy distributions of the two samples. In the data sample, aside from di-muon events in which the deposited energies of both tracks are low, there are also abnormal events where only one track has low energy deposition (marked by red boxes).
\label{fig:h2ee}}
\end{figure}

%\begin{figure}[htbp]
%\centering
%\includegraphics[width=0.45\textwidth]{lpnormph.eps}
%%\includegraphics[width=0.45\textwidth]{h2p1p2.eps} 
%\caption{
%Arbitrarily scaled distributions of normalized pulse heights of electrons and muons for tracks at around 2.1 GeV.
%\label{fig:normph}}
%\end{figure}

The integrated luminosity is calculated with the equation

\begin{equation}
\lum = \frac{N_{\rm obs}+N_{\rm cor}}{\sigma_{\rm
Bhabha} \times \eff},
\end{equation}
where $N_{\rm obs}$ is the number of observed Bhabha
events, $N_{\rm cor}$ is the number of events recovered by
the correction of the EMC readout error, $\sigma_{\rm Bhabha}$ is
the cross section of the Bhabha process, and $\eff$ is the
efficiency determined with the signal MC sample. The cross
sections are calculated by {\sc babayaga@nlo} generator using the
parameters listed in Table~\ref{tab:nlo_options}. All the input numbers
and the luminosity results are
listed in Table~\ref{tab:result_new}.

\section{Systematic Uncertainties}

The following sources of systematic uncertainties are considered:
the tracking efficiency, the requirements on the kinematic
variables, the limited sizes of the MC samples, the beam energy
measurement, the \revise{EMC readout correction}, and the MC
generator.

To estimate the systematic uncertainty of the tracking efficiency,
we employed an alternative selection criterion using
information from the EMC only. Here, at least two clusters
in the EMC are required; if more than two clusters are present, then the most energetic two are identified as the $\EE$ pair. The deposited energies of the two clusters are
required to be larger than \revise{$0.45 \times \ECM$}. The
polar angle of each cluster is required to satisfy
$|\cos\theta_{\rm EMC}|<0.8$. Additionally, $\Delta\phi$ is
required to be in the range of $[-40^\circ,~-5^\circ]$ or
$[5^\circ,~40^\circ]$, where $\Delta\phi=|\phi_1-\phi_2|-180^\circ$ and
$\phi_{1,2}$ are the azimuthal angles of the clusters in the EMC.
All the angles are boosted to the $\EE$ CM frame. The difference
between the luminosity obtained through this selection and the
original result before the \revise{EMC readout correction} is taken as the systematic
uncertainty arising from tracking efficiency.

The systematic uncertainties related to the requirements on the
kinematic variables are evaluated by varying the thresholds on the
variables. For the requirement on the EMC energy, the alternative
threshold is \revise{$ 0.41 \times \ECM$}; for the polar angle,
the alternative range is $[-0.75,~0.75]$; for the requirement on
momentum, the threshold is changed to \revise{$0.48\times \ECM$}.
For the invariant mass of the $e^+e^-$ pair, the
alternative threshold is 3.80~GeV/$c^2$, and the corresponding uncertainty is found to be negligible.

The statistical uncertainties of the MC samples size, each one having two million events with a selection efficiency of around 17\%, are estimated to be 0.2\% at each energy point.

%The uncertainty on the CM energy measurement is $\pm0.6$~MeV. To
%estimate its effect on the luminosity, we redo the analysis on the
%same data sets as if we were analyzing data at alternative energy
%points that are 0.6~MeV~\cite{BESIII:2020eyu} higher or lower and check the difference.
%No new MC sample is generated here to avoid the reintroduction of
%the statistical fluctuation with the MC samples, and instead we obtain the
%detection efficiencies and cross sections through the linear
%extrapolation from the nearby energy points.

The uncertainty on the CM energy measurement is $\pm0.6$~MeV \cite{BESIII:2020eyu};
its effect on the luminosity determination is estimated by repeating the analysis on the same data sets, while changing the CM energy value by plus and minus this value. To avoid an additional systematical uncertainty due to the MC data sample size, we obtain the detection efficiency and cross section values through the linear extrapolation from the nearby energy points. The uncertainty is estimated as the difference in integrated luminosity compared to  our standard result.  The small difference between the measured beam-energy spread~\cite{Yu:IPAC2016-TUYA01} and that used in the generation of the MC samples leads to a negligible bias in the analysis.

The systematic uncertainty from \revise{the EMC readout correction} is
estimated by comparing the results with an alternative \revise{correcting} method, where the events with one or two tracks not satisfying the energy requirements are selected, and the correction is estimated by fitting the two-dimensional d$E$/d$x$ distribution of the two tracks in these events with a model containing three components: Bhabha events, di-muon events, and a background of uniform distribution. The uncertainty is estimated as the difference in  result between the two correction methods.

The uncertainty on the predictions of the {\sc babayaga@nlo} generator is assigned to be 0.1\%, following  Ref.~\cite{Balossini:2008xr}.

The total uncertainty for each energy point is summarized in
Table~\ref{tab:error_new}. The uncertainties from different
sources are assumed to be independent, therefore the
total uncertainties are obtained by adding up the uncertainties in
quadrature.

\begin{table}[htbp]
\centering \caption{\label{tab:error_new} The relative systematic
uncertainties (in \%) for the integrated luminosities of the new XYZ data set.}
\begin{spacing}{1.29}
\renewcommand\tabcolsep{5.0pt}
\begin{threeparttable}
\begin{tabular}{c|c c c c c c c c}
\hline
Sample     (MeV)                 & 4190 & 4200 & 4210 & 4220 & 4237 & 4246 & 4270 & 4280 \\
\hline
Tracking efficiency         & 0.13 & 0.17 & 0.05 & 0.08 & 0.24 & 0.15 & 0.30 & 0.29 \\
Requirement on energy       & 0.12 & 0.11 & 0.12 & 0.14 & 0.09 & 0.12 & 0.13 & 0.10 \\
Requirement on $\cos\theta$ & 0.24 & 0.22 & 0.21 & 0.22 & 0.32 & 0.37 & 0.43 & 0.39 \\
Requirement on momentum     & 0.14 & 0.08 & 0.01 & 0.00 & 0.17 & 0.06 & 0.09 & 0.01 \\
%Requirement on $M(\EE)$     & 0.00 & 0.00 & 0.00 & 0.00 & 0.00 & 0.00 & 0.00 & 0.00 \\
CM energy     & 0.06 & 0.07 & 0.00 & 0.01 & 0.04 & 0.04 & 0.02 & 0.01 \\
MC sample size               & 0.2  & 0.2  & 0.2  & 0.2  & 0.2  & 0.2  & 0.2  & 0.2  \\
Correction of \revise{EMC readout error}   & 0.05 & 0.06 & 0.08 & 0.04 & 0.07 & 0.06 & 0.06 & 0.05 \\
Event generator             & 0.1  & 0.1  & 0.1  & 0.1  & 0.1  & 0.1  & 0.1  & 0.1  \\
\hline
Total                       & 0.41 & 0.39 & 0.35 & 0.35 & 0.45 & 0.50 & 0.59 & 0.55 \\
\hline
\end{tabular}
\end{threeparttable}
\end{spacing}
\end{table}

\section{Update on the Luminosity of the 2010--2014 Data Sets}

An update of the integrated luminosities for the 21 data samples collected in 2010-2014, previously reported in Ref.~\cite{PreviousAnalysis},  is needed in order to apply the \revise{EMC readout correction}, as described in the previous section, and to include additional events coming from the recovery of data files that were not originally available.  

The same procedure described for the 2016-2017 data samples has been applied for this update,
\revise{with} the same configurations of the MC generator, \revise{the same event selection
criteria,} and \revise{the same EMC readout correction}.  The {\sc babayaga@nlo} event
generator we use in this analysis has a significantly better
precision than the event generator used in the previous
analysis~\cite{PreviousAnalysis} (0.1\% versus 0.5\%), which
contributes to the reduction of the total uncertainties.

Table~\ref{tab:update_old_xyz} summarizes 
the updated integrated luminosities with statistical uncertainties, the
correction factors due to \revise{the EMC readout error}, and the comparison with
the results of the previous analysis. The results of the three lowest 
energy points (samples 3810, 3900, and 4009) are not updated,
since there is no file update and, according to the correlation between the amount of the \revise{EMC readout correction} and the CM energies in other data sets, the \revise{EMC readout correction} is expected
to be negligible at these energy points. We skip the unnecessary
update of these data samples to avoid dealing with the
computational difficulty of the MC sampling over narrow
resonances. For the remainder of the energy points, the updated results are $0.3\% - 4.8\%$ larger than the original values.
%with a contribution from the EMC readout correction that rises with energy.
\revise{The discrepancies are mainly caused by the EMC readout correction, and for a few data sets such as $4420_1$ and $4600$ there're also the contributions from recovery of data files.}
Figure ~\ref{fig:correction} shows the correlation between the sizes of \revise{the EMC readout correction} and the CM energies. 
\revise{This figure shows that the size of this correction grows exponentially as the CM energy increases, indicating that the problem may grow to be much more severe if BESIII is to operate in higher energy zones. Besides, the fact that the frequencies of EMC readout error fit well to an exponential model may hint at its mechanism.}

%As the figure suggests, the discrepancies between the updated luminosities and the original ones can be well explained by the contributions from the EMC correction. Also, the figure shows that the size of this correction grows exponentially as the CM energy increases.

\begin{table}[htbp]
\centering \caption{\label{tab:update_old_xyz}  The updated integrated 
luminosities of the 2010--2014 XYZ data sets, the correction factor due to the EMC readout error (\revise{$\sigma_{\rm EMC}=\rm{N}_{cor}/\rm{N}_{obs}$}),
and comparison with the previous results~\cite{PreviousAnalysis}. The first uncertainties are statistical and the second ones are systematic.}
\begin{tabular}{c c c c c}
\hline
 $E_{\rm cm} (\rm MeV)$ & $\sigma_{\rm EMC}$ (\%) & Updated $\lum$ ($\rm pb^{-1}$)
        & Previous $\lum$ ($\rm pb^{-1}$) & Difference (\%) \\
\hline
$3810$    & - & - & $50.54 \pm 0.03$ & - \\
$3900$    & - & - & $52.61 \pm 0.03 \pm 0.51$ & - \\
$4009$    & - & - & $482.0 \pm 0.1 \pm 4.7$ & -\\
$4090$    & \revise{$0.07 \pm 0.04$} &  $52.86 \pm 0.03 \pm 0.35 $ & $52.63 \pm 0.03 \pm 0.51$ & $+0.43$\\
$4190$    & \revise{$0.19 \pm 0.04$} &  $43.33 \pm 0.03 \pm 0.29$ & $43.09 \pm 0.03 \pm 0.42$ & $+0.56$\\
$4210$    & \revise{$0.24 \pm 0.00$} &  $54.95 \pm 0.03 \pm 0.36$ & $54.55 \pm 0.03 \pm 0.53$ & $+0.73$\\
$4220$    & \revise{$0.24 \pm 0.04$} &  $54.60 \pm 0.03 \pm 0.36$ & $54.13 \pm 0.03 \pm 0.53$ & $+0.86$\\
$4230_1$ & \revise{$0.27 \pm 0.04$} &  $44.54 \pm 0.03 \pm 0.29$ & $44.40 \pm 0.03 \pm 0.43$ & $+0.32$ \\
$4230_2$ & \revise{$0.27 \pm 0.04$} & $1056.4 \pm 0.1 \pm 7.0$ & $1047.3 \pm 0.1 \pm 10.1$ & $+0.86$ \\
$4245$     & \revise{$0.31 \pm 0.03$} &  $55.88 \pm 0.03 \pm 0.37$ & $55.59 \pm 0.04 \pm 0.54$ & $+0.53$\\
$4260_{1,2}$ & \revise{$0.34 \pm 0.04$} & $828.4 \pm 0.1 \pm 5.5$ & $523.7 \pm 0.1 \pm 5.1$ & $+0.32$\\
                                    & & & $302.0 \pm 0.1 \pm 3.0$ & \\
$4310$    & \revise{$0.51 \pm 0.06$} &  $45.08 \pm 0.03 \pm 0.30$ &  $44.90 \pm 0.03 \pm 0.44$ & $+0.40$ \\
$4360$    & \revise{$0.74 \pm 0.06$} & $544.0 \pm 0.1 \pm 3.6$ & $540.0 \pm 0.1 \pm 5.2$ & $+0.76$\\
$4390$    & \revise{$0.95 \pm 0.05$} &  $55.57 \pm 0.04 \pm 0.37$ & $55.18 \pm 0.04 \pm 0.54$ & $+0.70$ \\
$4420_1$ & \revise{$1.13 \pm 0.07$} & $46.80 \pm 0.03 \pm 0.31$ & $44.67 \pm 0.03 \pm 0.43$ & $+4.77$ \\
$4420_2$ & \revise{$1.20 \pm 0.06$} & $1043.9 \pm 0.1 \pm 6.9$ & $1028.9 \pm 0.1 \pm 10.0$ & $+1.45$\\
$4470$    & \revise{$1.71 \pm 0.03$} & $111.09 \pm 0.04 \pm 0.73$ & $109.94 \pm 0.04 \pm 1.07$ & $+1.05$ \\
$4530$    & \revise{$2.38 \pm 0.11$} & $112.12 \pm 0.04 \pm 0.73$ & $109.98 \pm 0.04 \pm 1.07$ & $+1.95$ \\
$4575$    & \revise{$3.13 \pm 0.14$} &  $48.93 \pm 0.03 \pm 0.32$ & $47.67 \pm 0.03 \pm 0.46$ & $+2.64$ \\
$4600$    & \revise{$3.51 \pm 0.15$} & $586.9 \pm 0.1 \pm 3.9$ & $566.9 \pm 0.1 \pm 5.5$ & $+3.52$ \\
\hline
\end{tabular}
\end{table}

\begin{figure}[htbp]
\centering
\includegraphics[width=0.75\textwidth]{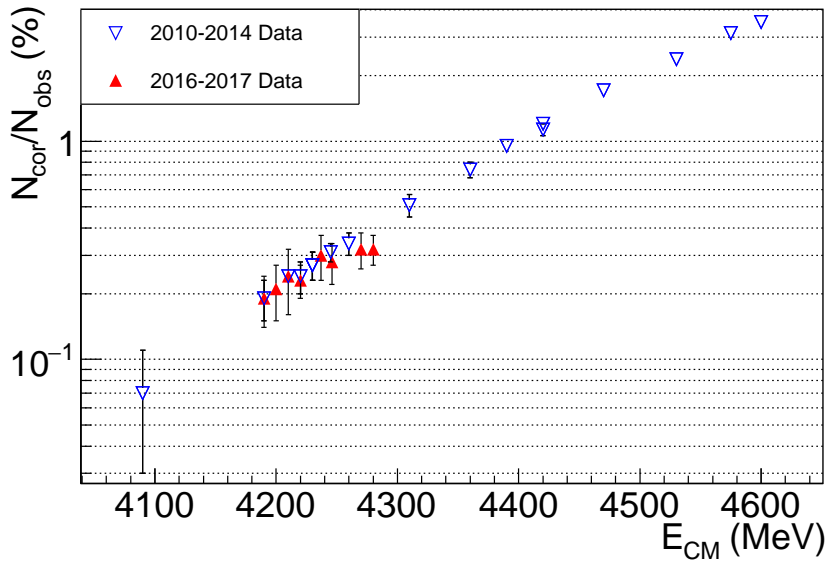}

\caption{
\oldchange{
The logarithm of relative sizes of \revise{the EMC readout correction} at different CM energies for 2010-2014 (blue) and 2016-2017 (red) XYZ data sets. In this plot, the data points have a seemingly linear relationship, suggesting that the  \revise{frequencies of the EMC readout errors} may grow exponentially as the CM energy further increases.}
\label{fig:correction}}
\end{figure}

Systematic uncertainties are assigned following the same procedure as for the 2016--2017 data sets.
For each source of uncertainty, we take the maximum uncertainty
for all the energy points.
These uncertainties from individual sources are then added up in
quadrature to obtain the total systematic uncertainty. The result
is summarized in Table~\ref{tab:error_old_xyz}. The total
systematic uncertainty is determined to be 0.66\% for all the
energy points except for the lowest three energy points, for which we
quote the original uncertainty 0.97\%~\cite{PreviousAnalysis}.

\begin{table}[htbp]
\centering \caption{\label{tab:error_old_xyz} The systematic
uncertainties of the integrated luminosities of the 2010--2014 XYZ
data samples, excluding those at the lowest three energy points.
\revise{For each source of uncertainty, the maximum value across all the energy points is taken as the overall estimation.}}
\begin{tabular}{c c}
\hline
 Source                          &  Relative uncertainty (\%)  \\\hline
Tracking efficiency              & $0.42$ \\
Requirement on energy            & $0.28$ \\
Requirement on $\cos\theta$    & $0.14$ \\
Requirement on momentum          & $0.29$ \\
CM energy                      & $0.07$ \\
%{\sc babayaga@nlo} generator     & $0.1$  \\
MC sample size                    & $0.20$  \\
Correction of \revise{EMC readout errors}        & $0.15$ \\ 
Event generator     & $0.10$  \\ \hline
Total                            & $0.66$ \\
\hline
\end{tabular}
\end{table}

%-----------------------------------------------------------------------------------------------------------------------------------------------------------------------------
\section{Summary}

We have measured the integrated luminosities of the XYZ data sets taken
at BESIII from 2016 to 2017, and the results
are listed in Table~\ref{tab:result_new}.
Additionally, we have updated the luminosity measurement
of the XYZ data taken from 2010 to 2014, with corrections arising from an improved understanding of the  EMC performance and the recovery of data files, as shown in Table~\ref{tab:update_old_xyz}. 
These high precision results are of fundamental importance for the measurements of the production cross
sections of the XYZ particles as well as those of  conventional charmonium states  in this energy range, which will enable a more precise comparison with the predictions  of the quark model and an improved  understanding of QCD. The results presented in this work have been used in several recent analyses of the BESIII collaboration (e.g. see
Refs.~\cite{r6,r10,r4}) and will be used by many other analyses in the future.

\oldchange{Figure ~\ref{fig:correction} shows that the impact of the \revise{EMC readout errors} may grow exponentially to above 10\% as the CM energy increases to around 5.0 GeV. This means that the dangerous effect warrants more inspection if BESIII is to operate in higher energy zones with BEPCII update project in the future.}

%--------------------------------------------------------------------------------------------------

\section{Acknowledgement}
The BESIII collaboration thanks the staff of BEPCII and the IHEP computing center for their strong support. This work is supported in part by National Key R\&D Program of China under Contracts Nos. 2020YFA0406300, 2020YFA0406400; National Natural Science Foundation of China (NSFC) under Contracts Nos. 11625523, 11635010, 11735014, 11822506, 11835012, 11935015, 11935016, 11935018, 11961141012, 12022510, 12025502, 12035009, 12035013, 12061131003; the Chinese Academy of Sciences (CAS) Large-Scale Scientific Facility Program; Joint Large-Scale Scientific Facility Funds of the NSFC and CAS under Contracts Nos. U1732263, U1832207; CAS Key Research Program of Frontier Sciences under Contract No. QYZDJ-SSW-SLH040; 100 Talents Program of CAS; INPAC and Shanghai Key Laboratory for Particle Physics and Cosmology; ERC under Contract No. 758462; European Union Horizon 2020 research and innovation programme under Contract No. Marie Sklodowska-Curie grant agreement No 894790; German Research Foundation DFG under Contracts Nos. 443159800, Collaborative Research Center CRC 1044, GRK 2149; Istituto Nazionale di Fisica Nucleare, Italy; Ministry of Development of Turkey under Contract No. DPT2006K-120470; National Science and Technology fund; Olle Engkvist Foundation under Contract No. 200-0605; STFC (United Kingdom); The Knut and Alice Wallenberg Foundation (Sweden) under Contract No. 2016.0157; The Royal Society, UK under Contracts Nos. DH140054, DH160214; The Swedish Research Council; U. S. Department of Energy under Contracts Nos. DE-FG02-05ER41374, DE-SC-0012069.

\end{document}